\documentclass[sigconf,screen]{acmart}

\AtBeginDocument{%
  }

\setcopyright{acmlicensed}
\copyrightyear{2026}
\acmYear{2026}
\acmDOI{XXXXXXX.XXXXXXX} 
\acmConference[FSE 2026]{ACM International Conference on the Foundations of Software Engineering}{July 05--09, 2026}{Montreal, Canada}

\acmBooktitle{Companion Proceedings of the 34th ACM Symposium on the Foundations of Software Engineering (FSE '26), June 5--9, 2026, Montreal, Canada}
\acmISBN{c978-1-4503-XXXX-X/2018/06}

\usepackage[table]{xcolor} 
\usepackage{tikz}
\usepackage{multirow}
\usepackage{diagbox}
\usetikzlibrary{positioning, fit, shapes}

\usepackage{tablefootnote}
\usepackage{listings}
\usepackage[edges]{forest}
\usetikzlibrary{positioning}
\usepackage[most]{tcolorbox}

\definecolor{purplish}{HTML}{D8DFE3}
\definecolor{purplishlight}{HTML}{EBEFF3}
\definecolor{purplishdark}{HTML}{009e73}

\newtcolorbox[auto counter,number within=section]{rqbox}[2]{
    nameref=#1,
    title=\small{#1}, 
    enhanced,
    attach boxed title to top left={yshift=-6pt, xshift=8pt},
    boxed title style={size=small,boxsep=1pt},
    colframe=purplishdark,colback=white,colbacktitle=purplishdark,
    boxsep=2pt,left=2pt,right=2pt,top=6pt,bottom=2pt,middle=2pt
}

\newcommand{\rqtextone}{\textbf{Research Question --} What is the relationship between team diversity -- specifically in terms of sexual orientation -- and fairness awareness in software development tasks?}

\begin{document}

\title{Team Diversity Promotes Software Fairness: An Experiment on Fairness-Aware Requirements Prioritization}

\author{Cleyton Magalhães}
\affiliation{
  \institution{UFRPE}
  \city{Recife}
  \country{Brazil}
}
\email{cleyton.vanut@ufrpe.br}

\author{Ronnie de Souza Santos}
\affiliation{
  \institution{University of Calgary}
  \city{Calgary}
  \state{Alberta}
  \country{Canada}
}
\email{ronnie.desouzasantos@ucalgary.ca}

\author{Bimpe Ayoola}
\affiliation{
  \institution{Dalhousie University}
  \city{Halifax}
  \state{Nova Scotia}
  \country{Canada}
}
\email{bimpe.ayoola@dal.ca}

\author{Brody Stuart-Verner}
\affiliation{
  \institution{University of Calgary}
  \city{Calgary}
  \state{Alberta}
  \country{Canada}
}
\email{brody.stuartverner@ucalgary.ca}

\author{Italo Santos}
\affiliation{
  \institution{University of Hawaii at Manoa}
  \city{Honolulu}
  \state{Hawaii}
  \country{USA}
}
\email{isantos3@hawaii.edu}

\begin{abstract}
\textbf{Background:} Fairness and diversity are receiving growing attention in software engineering, particularly as AI and machine learning systems increasingly influence decision-making processes. While fairness is often examined at the algorithmic or data level, there is limited understanding of how it is addressed during the early stages of software development. Moreover, little is known about how team diversity affects fairness-related decisions in software projects. \textbf{Aims:} This study investigates how diversity in software teams influences fairness-aware behavior during requirements prioritization. \textbf{Method:} A controlled experiment was conducted with 27 pairs of software engineering students, including 13 LGBTQ-diverse pairs and 14 non-diverse pairs. Each pair prioritized user stories with varying fairness implications. Descriptive statistics were used to analyze attitudes and prioritization outcomes, and thematic analysis was applied to examine the reasoning behind participants’ decisions.  \textbf{Results:} Both groups demonstrated general alignment with fairness principles, prioritizing features that promoted equitable treatment and rejecting those that posed fairness risks. However, LGBTQ-diverse pairs were more consistent in rejecting fairness-risking stories and made fewer fairness-related misprioritization errors. Their reasoning emphasized inclusion, non-discrimination, and ethical responsibility, whereas non-diverse pairs adopted a more pragmatic, goal-oriented perspective. \textbf{Conclusions:} The findings indicate that fairness should be considered from the earliest stages of software development. Team diversity can enhance the identification and interpretation of fairness issues during requirements analysis, fostering more reflective and inclusive decision-making.

\end{abstract}




\maketitle

\section{Introduction} \label{sec:intro}

Persistent disparities in representation remain a significant challenge in software engineering, with women, racialized individuals, people with disabilities, and LGBTQIA+ communities being particularly underrepresented in academic and professional domains~\cite{albusays2021diversity, rodriguez2021perceived, aleem2023practicing}. These disparities often begin in academia due to unwelcoming cultures and institutional norms that marginalize underrepresented students~\cite{de2023diversity, oliveira2024navigating, gama2024s} and extend to industry, where equity, diversity and inclusion (EDI) efforts are frequently superficial or disconnected from engineering practices~\cite{albusays2021diversity, menezes2018diversity, rodriguez2021perceived, aleem2023practicing}. Despite these challenges, diverse software teams have been shown to enhance creativity, collaboration, and problem-solving by incorporating a wider range of perspectives and addressing more inclusive user needs~\cite{albusays2021diversity, rodriguez2021perceived, menezes2018diversity, aleem2023practicing}. These benefits are especially pronounced when supported by inclusive environments that promote autonomy, empathy, and shared competence~\cite{devathasan2025empathy}.

In recent years, concerns about fairness, bias, and the broader social consequences of software systems have become the focus of many discussions, especially systems powered by artificial intelligence (AI)~\cite{kordzadeh2022algorithmic, dehal2024exposing, ferrer2021bias}. Software that embeds biased assumptions can amplify existing inequalities, marginalize vulnerable groups, and erode public trust in technology. These risks are particularly problematic in contexts where algorithmic decisions can affect access to essential services, such as healthcare, education, finance, and justice~\cite{desoftware, kordzadeh2022algorithmic, dehal2024exposing}.
Addressing fairness in software development has, therefore, become a key concern in the software industry~\cite{brun2018software}, requiring not only technical solutions but also a broader awareness of the social dynamics that influence how AI systems are designed, developed, and deployed, including the perspectives represented within development teams~\cite{adams2020diversity, desoftware}.

In this scenario, to ensure that fairness considerations are incorporated throughout the software development process, team diversity arises as an important strategy~\cite{rostcheck2025elephant, adams2020diversity, fosch2022diversity, de2022picture}. Broader lived experiences enable teams to detect potential harms earlier and design systems that better reflect societal needs~\cite{rostcheck2025elephant, desoftware, de2022picture}. Previous works showed that teams composed of individuals with varied backgrounds are more likely to anticipate the different ways software can affect communities and challenge assumptions that might otherwise go unnoticed~\cite{adams2020diversity, rostcheck2025elephant}. In particular, in AI development, where transparency and fairness are key aspects, the absence of such perspectives has been recognized as a factor that diminishes fairness awareness and reinforces dominant biases~\cite{adams2020diversity, fosch2022diversity}. However, while the value of diversity in mitigating bias is often asserted, empirical evidence on how it concretely influences fairness-oriented reasoning during software development remains scarce. Most existing studies discuss demographic representation or inclusive practices at organizational levels but rarely examine diversity as a determinant of fairness-aware decision-making in technical activities. To our knowledge, no prior empirical work has investigated how diversity in sexual orientation shapes fairness awareness in software requirements analysis or prioritization. This study addresses that gap by providing one of the first controlled investigations into the behavioral link between LGBTQIA+ team composition and fairness-aware software reasoning.

Considering that fairness in AI systems is shaped not only by technical choices but also by the social composition of development teams~\cite{adams2020diversity, rostcheck2025elephant, de2022picture, desoftware}, in this study, our main goal is to investigate the relationship between team diversity and fairness awareness in software engineering. Specifically, we conducted an experiment to investigate whether diverse teams are more likely to recognize, prioritize, and address fairness-related concerns when analyzing and prioritizing software requirements. The following research question guided our research:

 \newcommand{\rqone}[2][]{
     \begin{rqbox}{\textbf{Research Question}}{#2}
         \rqtextone
         #1
     \end{rqbox}
 }

 \rqone{}

To answer our research question, we conducted an experiment with software engineering students working in pairs. The pairs' work was analyzed based on sexual orientation: non-diverse or homogeneous pairs (both participants identified as heterosexual) and LGBTQ-diverse pairs (at least one participant identified as LGBTQIA+). Each pair evaluated a set of software requirements, identifying and prioritizing fairness-related features and deprioritizing those that could lead to discriminatory outcomes. This setup allowed us to explore how team composition might influence fairness awareness during the early stages of software development, with a particular focus on team diversity in terms of sexual orientation. Overall, our paper makes three primary contributions. First, it operationalizes fairness-aware attitudes in the context of software requirements. Second, it provides empirical evidence on how team diversity influences fairness-oriented decision-making in software projects. Third, it examines the contribution of LGBTQIA+ software professionals to fairness-aware development, showing how their participation supports ethical and inclusive practices in software engineering.

Our study is relevant to ongoing discussions about fairness in AI systems, where many biases originate from unexplored requirements rather than from models themselves~\cite{baresi2023understanding}. By observing how diverse teams identify and reason about fairness risks early in development, this work demonstrates how diversity can help prevent biased requirements from evolving into discriminatory system behaviors. Finally, our findings connect equity, diversity, and inclusion to improvements in software quality and social responsibility. In a period when the role of EDI in the technology sector is being questioned, this study provides evidence that inclusive team composition contributes directly to the fairness and reliability of AI-enabled systems.

From this introduction, this study is organized as follows. In Section~\ref{sec:background}, we present a literature review on software fairness and team diversity. Section~\ref{sec:method} describes our experiment. In Section~\ref{sec:findings}, we present our findings, which are discussed in Section~\ref{sec:discussion}, along with the implications and limitations of this study. Finally, Section~\ref{sec:conclusion} summarizes our contributions and final considerations.

\section{Background and Related Work}  \label{sec:background}

This section elaborates on key concepts that underpin this research: team diversity and software fairness, outlining their relevance, current understanding, and the gaps this study seeks to address.

\subsection{Software Team Diversity}
In software engineering, diversity refers to the range of individual characteristics and experiences that developers bring to a team, such as gender, race, age, nationality, and professional background~\cite{rodriguez2021perceived, albusays2021diversity}. These attributes can influence how individuals are treated and how they collaborate within development environments~\cite{catolino2019gender, verwijs2023double}. Recent studies conceptualize diversity not only as a demographic attribute but as a dynamic factor that shapes communication, coordination, and psychological safety in software teams~\cite{mason2024diversity, kohl2022benefits, de2023benefits}. Diversity can affect role distribution, decision-making, and the extent to which developers feel safe contributing ideas~\cite{vasilescu2015gender, rodriguez2021perceived, ortu2017diverse}. As software development increasingly depends on collaboration, particularly in distributed and open-source contexts, understanding how diverse identities influence team dynamics has become a significant area of inquiry~\cite{verwijs2023double, ortu2017diverse, rodriguez2021perceived}.

Despite growing attention, research on diversity in software engineering remains uneven, with gender receiving the most focus~\cite{catolino2019gender, vasilescu2015gender, kohl2022benefits, boman2024breaking}. Studies associate gender-diverse teams with improved innovation and communication~\cite{catolino2019gender, vasilescu2015gender}, while cultural and national diversity have been linked to engagement and language negotiation in distributed teams~\cite{ortu2017diverse, raithel2021team}. Age and tenure diversity are less examined but appear related to knowledge sharing and self-efficacy~\cite{vasilescu2015gender, van2023still}. Race remains largely overlooked~\cite{mason2024diversity}, and research on sexual orientation has only recently emerged, examining the inclusion of LGBTQIA+ professionals~\cite{boman2024breaking, de2023benefits}. Intersectional analyses that account for overlapping identities, such as race and gender, remain scarce~\cite{rodriguez2021perceived, shameer2023relationship}.

Diversity’s effects are contingent on team culture. Inclusive environments enhance collaboration, creativity, and equity~\cite{mason2024diversity, catolino2019gender, vasilescu2015gender}, particularly where psychological safety allows open dialogue~\cite{verwijs2023double}. Conversely, exclusionary settings can produce anxiety, communication barriers, and attrition among underrepresented developers~\cite{kohl2022benefits, hankins2023does, kim2018differential}. These dynamics indicate that diversity alone does not guarantee equitable outcomes; it must be accompanied by structures that foster trust, fairness, and participation. Such findings underscore that who participates in development influences how fairness, inclusion, and responsibility are embedded in software systems.

\subsection{Software Fairness}
Building on this connection, software fairness concerns the ability of systems to provide equitable treatment and outcomes to all users, regardless of their identity or background~\cite{galhotra2017fairness, brun2018software, sesari2024understanding}. As AI-enabled applications increasingly mediate access to opportunities and services, ensuring fairness has become a critical software quality attribute. Fairness issues often arise not from functional failures but from implicit biases in data, requirements, or design assumptions~\cite{galhotra2017fairness, chen2024fairness, brun2018software}. Because such issues may remain invisible in standard testing, systems can appear to function correctly while producing unjust outcomes for particular groups~\cite{chen2024fairness, desoftware}. This has led to growing recognition of fairness as a non-functional requirement that must be addressed throughout the development lifecycle~\cite{brun2018software, zhang2021ignorance, chen2024fairness, sesari2024understanding}.

Unaddressed fairness problems can reinforce structural inequalities, excluding or disadvantaging certain populations~\cite{galhotra2017fairness, chen2024fairness, desoftware}. Fairness requirements define expectations for preventing prejudice or favoritism based on sensitive characteristics, framing fairness as a measurable design objective~\cite{baresi2023understanding}. However, specifying and verifying fairness requirements remains complex because fairness depends on context, data, and normative interpretation~\cite{baresi2023understanding, villamizar2021requirements}. Real-world failures illustrate these challenges, such as predictive policing tools that overestimate risk for Black individuals or hiring algorithms that penalize women due to biased datasets~\cite{ferrer2021bias, dehal2024exposing, desoftware}. These examples reveal that fairness violations not only harm individuals but also undermine the credibility of AI systems, emphasizing the need for proactive rather than reactive fairness practices~\cite{kordzadeh2022algorithmic, chen2024fairness}.

Addressing fairness is therefore a socio-technical process shaped by who participates in development and how decisions are made~\cite{desoftware}. Biases often emerge during early-stage tasks such as requirements prioritization or design trade-offs, reflecting the dominant perspectives within teams~\cite{zhang2021ignorance, desoftware}. Teams that include diverse perspectives are better equipped to identify risks, question assumptions, and prevent the embedding of harmful biases~\cite{adams2020diversity}. In this sense, fairness is not only a property of software artifacts but also a function of the social composition and practices of development teams.

\subsection{Fairness Requirements}

Fairness requirements specify the conditions under which software systems make equitable decisions and avoid discriminatory outcomes across user groups~\cite{gjorgjevikj2023requirements, chen2024fairness}. They extend traditional non-functional requirements by embedding ethical and social constraints into data processing and decision-making~\cite{brun2018software}. Prior research conceptualizes fairness as a quality attribute that should be elicited and analyzed during requirements engineering rather than verified post hoc~\cite{brun2018software, ramadan2025towards}. These requirements are particularly salient in AI-enabled contexts such as finance, recruitment, and healthcare, where unfair outputs can produce significant social harm~\cite{chen2024fairness, lavalle2022law}. Because fairness is context dependent, multiple interpretations coexist, such as equality of opportunity, demographic parity, and individual fairness, each reflecting distinct normative assumptions and design implications~\cite{ramadan2025towards}.  

The elicitation of fairness requirements involves identifying potential sources of bias in data, models, and decision processes~\cite{gjorgjevikj2023requirements}. Legal-oriented approaches align software requirements with duties derived from regulation and policy, mapping protected attributes to system components through legal modeling frameworks~\cite{lavalle2022law}. Other approaches emphasize participatory analysis with domain experts and affected stakeholders to translate ethical principles into verifiable specifications~\cite{gjorgjevikj2023requirements, chen2024fairness}. Techniques such as fairness checklists, domain taxonomies, and scenario-based evaluation help operationalize fairness as a measurable quality within the software lifecycle~\cite{ramadan2025towards, chen2024fairness}.  

Despite these advances, the treatment of fairness in requirements engineering remains fragmented and inconsistent~\cite{ramadan2025towards}. Fairness knowledge is distributed across legal, ethical, and technical domains, leading to underspecified or unverifiable requirements~\cite{chen2024fairness}. Practitioners often disagree on which fairness notions apply or how to balance fairness against other system goals such as accuracy or efficiency~\cite{gjorgjevikj2023requirements, brun2018software}. Contextual variability further complicates analysis, as fairness definitions appropriate for one domain or dataset may not generalize elsewhere~\cite{lavalle2022law}. Translating ethical or legal norms into technical constraints introduces interpretation bias, highlighting the need for integrative frameworks that connect legal reasoning, domain expertise, and empirical validation~\cite{ramadan2025towards, brun2018software}.
\section{Method} \label{sec:method}
This study employs an experimental approach to investigate the relationship between team diversity and fairness awareness during software requirements analysis. This section describes the design, instruments, recruitment, data collection, and analysis procedures, emphasizing methodological rigor, ethical safeguards, and transparency.

\subsection{Study Design}  
\label{sec:design}

This exploratory study examines whether diversity, specifically in terms of sexual orientation, affects fairness awareness during software requirements analysis. Rather than testing predefined hypotheses, this study was guided by a general research question presented in Section~\ref{sec:intro}. The design drew on previous research that used persona models and stakeholder maps to investigate whether design interventions could influence how software engineering students prioritize socially responsible and antisocial features in a backlog~\cite{ayoola2024user}. While that study focused on external tools that promote prosocial decision making, our work investigates whether team diversity shapes fairness-aware decision making during requirements prioritization. We followed established methodological guidelines for experiments involving human participants in software engineering~\cite{jedlitschka2008reporting, falessi2018empirical, ralph2020empirical}. Our procedure followed the steps below.

\begin{enumerate}
    \item \textbf{Participant preparation.}  
    Participants were student volunteers enrolled in a project-based course that applied agile practices. All had received formal instruction in requirements engineering through their coursework. Participation was voluntary and unrelated to course assessment.

    \item \textbf{Pilot construction of user stories.}  
    Before the study, a pilot phase was conducted with three researchers who were not part of the author team. The pilot produced a stable list of twenty-four unambiguous user stories labeled as fair, unfair, or neutral. These classifications were known to the researchers but not to the students.

    \item \textbf{Pair formation.}  
    During the study sessions, participants were paired randomly in class before any demographic disclosure. This ensured that pairing was not based on self-identified attributes to ensure anonymity.

    \item \textbf{Task briefing and instructions.}  
    At the start of the activity, each pair received the list of user stories and written instructions (example available on Table \ref{tab:userstory}). Pairs were told to discuss each story collaboratively, assign a priority score from one (low) to four (high), and provide a brief justification for their decisions. When both agreed that a story should be implemented immediately, they assigned the highest priority. When they identified a story as potentially unfair or discriminatory, they reduced or removed its priority. In a real-world context, this would represent postponing implementation until further clarification was obtained.

    \item \textbf{Execution of the task.}  
    Each pair worked together for approximately one hour. All decisions were made jointly, ensuring that prioritization reflected discussion rather than individual preferences.

    \item \textbf{Collection of demographic data.}  
    After completing the prioritization task, each participant filled out an individual demographic questionnaire privately. This form included optional self-identifications for gender, race, sexual orientation, disability, neurodivergence, and socioeconomic background. No names or identifiers were requested, and participants could skip any category. Forms were not shared between pair members to guarantee that their personal characteristics remain undisclosed.

    \item \textbf{Anonymization and coding.}  
    Once both members of a pair delivered their materials, the prioritization sheet and demographic forms were collected and shuffled to remove any ordering pattern that could link data to individuals or sessions. Each set was assigned a numeric code representing a pair, allowing analysis at the pair level while maintaining anonymity among the individuals and pairs.

    \item \textbf{Post-hoc demographic verification.}  
   After all sessions were completed, anonymized self-identified demographic data were collected to retrospectively confirm the composition of participant pairs. This procedure allowed us to determine which diversity dimensions, such as gender, race, or sexual orientation, could be meaningfully analyzed. Among these, sexual orientation emerged as the most suitable focus for this study, as the distribution of participants resulted in a comparable number of LGBTQ-diverse and non-diverse pairs. Analyzing outcomes by sexual orientation after the task introduces potential validity threats, which are discussed in Section~\ref{sec:validity}. However, collecting this attribute in advance was both ethically and practically challenging, since LGBTQ professionals constitute a hidden population and disclosure of identity cannot be expected upon request \cite{de2024hidden}. The decision to collect this information post-task therefore reflected a deliberate trade-off between ethical responsibility and analytical feasibility.

    \item \textbf{Data analysis.}  
    Quantitative and qualitative analyses were conducted to examine how diversity influenced the recognition, prioritization, and discussion of fairness-related concerns. The analytical procedures are described in Section~\ref{sec:analysis}.
\end{enumerate}

This structured design preserved ethical safeguards while enabling a comparative investigation of team diversity and fairness awareness. Some limitations accompany this approach. Post-hoc classification based on self-reported data may have introduced confounding variables, as diversity attributes were unevenly distributed. The use of students constrains external validity, since their professional experience may differ from that of practitioners. Finally, voluntary participation may have led to self-selection bias, as students interested in fairness could have been more likely to participate. These limitations are common to exploratory studies involving human participants but were mitigated through transparency, ethical oversight, and the triangulation of quantitative and qualitative evidence. We discuss this further in Section \ref{sec:validity}

\subsection{Instrumentation}  
\label{sec:instrument}

The experimental instrument was designed to simulate a realistic requirements analysis task while allowing the observation of fairness-aware decision making. We drew conceptually on previous research that employed categorically balanced sets of user stories to study how design interventions influence the prioritization of socially responsible and antisocial software features~\cite{ayoola2024user}. Building on this approach, we developed a fictional case involving a financial institution that planned to deploy an AI based system to support credit risk evaluation and loan approval. The scenario was selected because fairness considerations are particularly relevant in automated financial decision making, where predictive analytics may reproduce or amplify social bias. \textbf{Construction of user stories.}  
An initial pool of user stories was drafted to represent a range of fairness implications associated with system design. Each story described a potential feature of the AI based financing system and was worded to resemble items commonly found in agile backlogs. To establish a balanced dataset, the stories were classified into three categories: (1) fairness aligned, referring to features that promote equitable treatment; (2) fairness risking, referring to features that could introduce bias or exclusion; and (3) neutral, referring to features unrelated to fairness. To ensure the reliability of this classification, a pilot study was conducted before the experiment. Three reviewers with distinct disciplinary backgrounds, one in software engineering, one in human computer interaction, and one in the social sciences, independently classified each story. The reviewers discussed disagreements and refined the wording or category assignment until full consensus was achieved. After three iterations, this process produced a validated and balanced set of twenty four user stories: eight fairness aligned, eight fairness risking, and eight neutral. Two additional stories served as attention checks and were excluded from the analysis. Table~\ref{tab:userstory} presents examples of the validated stories, which are fully available in our replication package. \textbf{Supporting materials.}  
To capture participants’ reasoning, the prioritization sheet included an open comment field beside each story, where pairs could briefly explain the rationale for their decisions. A post task reflective question invited each pair to summarize how they approached the prioritization exercise, the factors they considered most relevant, and how they reached consensus. A separate demographic questionnaire allowed participants to voluntarily and anonymously self identify across several dimensions of diversity, including gender, race, sexual orientation, disability, neurodivergence, and socioeconomic background. These instruments were designed to collect sufficient contextual and reflective information while maintaining participant anonymity and ethical compliance.

\begin{table}[h]
\caption{User Story Prioritization Example}
\label{tab:userstory}
\centering
\scriptsize
\begin{tabular}{p{7cm}}
\toprule
\textbf{User Story} \\
\midrule
\textit{US-1: As a loan officer, I want to apply risk assessment algorithms to flag applicants from low-income neighborhoods for additional scrutiny to minimize risk.} \\
\vspace{0.1cm}
\textbf{Priority (Select 1–4):} \\
4 – High Priority \\
3 – Medium Priority \\
2 – Low Priority \\
1 – Very Low / Do Not Implement \\
\vspace{0.1cm}
\\ \hline
\textit{US-2: As a loan officer, I want the system to automate document verification and approval workflows to reduce processing time and improve efficiency.} \\
\vspace{0.1cm}
\textbf{Priority (Select 1–4):} \\
4 – High Priority \\
3 – Medium Priority \\
2 – Low Priority \\
1 – Very Low / Do Not Implement \\
\vspace{0.1cm}
\\ \hline
\textit{US-3: As a customer with limited credit history, I want the software to consider alternative data points for assessing my creditworthiness.} \\
\vspace{0.1cm}
\textbf{Priority (Select 1–4):} \\
4 – High Priority \\
3 – Medium Priority \\
2 – Low Priority \\
1 – Very Low / Do Not Implement \\
\bottomrule
\end{tabular}
\end{table}

\subsection{Recruitment}  
\label{sec:recruitment}

Participants were recruited from undergraduate programs in software engineering and computer science. Eligibility required prior exposure to requirements engineering and teamwork to ensure an informed understanding of the activity. Each participant had completed an introductory course on software requirements and was enrolled in, or had completed, an advanced course that involved project based learning and agile practices. This ensured a shared foundation of technical preparation and collaborative experience. To approximate professional conditions, students with prior industry exposure through internships or part time roles were encouraged to participate. This combination of academic and applied experience provided a realistic yet controlled context for investigating fairness awareness during requirements analysis. Participation was voluntary and conducted outside regular course activities. Two sessions were organized to accommodate participants’ availability. A total of fifty four students took part, forming twenty seven pairs. Eighteen pairs participated in the first session and nine in the second. Each session included a mix of pairs with different diversity compositions who completed the same task under equivalent conditions. The study received ethics approval, and informed consent was obtained from all participants before data collection.

\subsection{Data Collection}  
\label{sec:datacollection}

Each pair received the prioritization task during an in person session and worked collaboratively to review the twenty four user stories. They assigned a priority score from one to four to each story and wrote brief justifications for their decisions. The task simulated a realistic requirements analysis exercise and was designed to prompt reflection on fairness, risk, and bias in software features.

After completing the task, each participant filled out a demographic form privately. The form collected optional self identifications across several diversity dimensions, including gender, race, sexual orientation, disability, neurodivergence, and socioeconomic background, but contained no names or identifying information. Pair members did not have access to one another’s demographic data. Once both members submitted their materials, the prioritization sheet and demographic forms were collected together and shuffled to remove any session specific ordering. Each set of documents was then assigned a numeric code representing a pair, allowing analysis at the pair level while protecting individual identities.

This process maintained strict anonymity. Because demographic information was collected after the activity, it was not possible to link individual participants to specific prioritization ratings. This approach preserved confidentiality but prevented follow up clarification on individual responses. The procedure followed approved ethics protocols, and all participants were informed of these safeguards during the consent process.

\subsection{Data Analysis}  
\label{sec:analysis}

The analysis combined quantitative and qualitative methods to explore how diversity influenced fairness awareness during requirements prioritization. The process involved two main stages: the descriptive statistical analysis of prioritization behaviors and the thematic interpretation of participants’ written justifications. \\

\noindent \textbf{Quantitative analysis.}  
All prioritization scores were first organized by pair and by story category: fairness aligned, fairness risking, or neutral. Each pair’s scores were then mapped to specific behavioral categories reflecting alignment with fairness objectives usig descriptive statistics~\cite{george2018descriptive}. The following behaviors were defined: (1) prioritizing fairness aligned stories, (2) deprioritizing fairness risking stories, (3) showing cautious support for fairness aligned stories (ratings of 2 or 3), and (4) showing cautious rejection of fairness risking stories (ratings of 2 or 3). Two composite indicators were also created: a \textit{pro fairness score}, combining the number of correct fairness supporting decisions (prioritizing aligned stories and deprioritizing risking stories), and a \textit{misprioritization score}, combining fairness inconsistent decisions (rating a fairness aligned story as 1 or a fairness risking story as 4). To compare groups of different sizes, we applied per pair normalization~\cite{martin2012eliciting, curran2013explorations}. Each pair’s behavioral counts were first computed individually and then averaged across the group. This approach ensured that every pair contributed equally to the group average, avoiding distortions due to the number of pairs or total ratings. For each behavior, we computed both the total count and the per pair average. When relevant, composite indicators were calculated by summing correct or incorrect behaviors, following methods used in behavioral and accuracy based research.  This procedure enabled systematic comparison between the LGBTQIA+ diverse group (13 pairs) and the non diverse group (14 pairs) across six fairness related behaviors: (1) prioritizing fairness aligned stories, (2) deprioritizing fairness risking stories, (3) pro fairness behavior, (4) misprioritization, (5) caution with fairness aligned stories, and (6) caution with fairness risking stories. Descriptive results were interpreted as indicative of behavioral trends rather than statistical generalizations, since inferential testing was not suitable for the sample size. Interpretations focused on how observed differences suggested variations in fairness reasoning and caution across groups. \\

\begin{figure}[!ht]
    \centering
    \includegraphics[width=9cm]{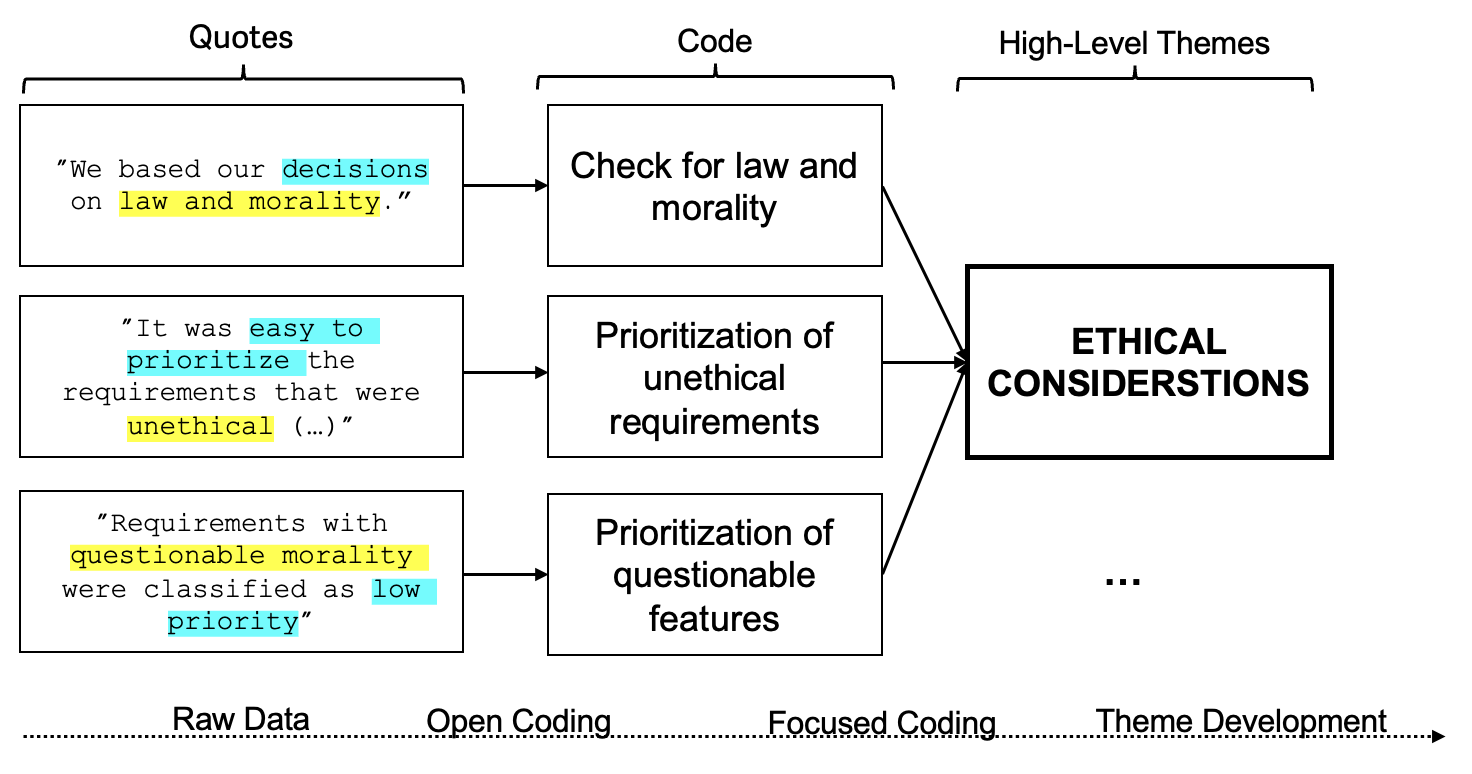}
    \vspace{-1em}
    \caption{Thematic Analysis}
    \vspace{-1em}
    \label{fig:quali}
\end{figure}

\noindent \textbf{Qualitative analysis.}  
Participants’ written justifications were analyzed to contextualize quantitative patterns and to identify the reasoning behind prioritization choices. The analysis followed a structured three step procedure consistent with established thematic analysis practices~\cite{cruzes2011recommended}, as illustrated in Figure \ref{fig:quali}. In the first step, \textit{open coding}, each written response was examined to capture the main idea or consideration expressed by participants. Codes reflected the reasoning underlying each justification, such as moral judgment, inclusivity, or perceived technical importance. In the second step, \textit{focused coding}, similar open codes were grouped into broader conceptual categories that represented recurring reasoning patterns, such as ethical reflection, attention to non discrimination, or awareness of project scope. In the third step, \textit{theme building}, the focused codes were synthesized into higher level themes aligned with the research objective of understanding fairness awareness in requirements decision making. The resulting themes were Inclusion, Non Discrimination, Ethical Considerations, Project Scope and Value, Technical Aspects, Legal and Compliance, and Social Business Balance. Two researchers independently coded the data, compared interpretations, and resolved disagreements through discussion. This collaborative process enhanced reliability and reduced subjective bias. Thematic findings were then integrated with descriptive patterns to provide a comprehensive understanding of how diversity, particularly LGBTQIA+ representation, influenced fairness reasoning, caution, and prioritization behavior.

\subsection{Ethics} \label{sec:ethics}
The study was reviewed and approved by the institutional research ethics board. All participants provided informed consent before participation. They were informed of the study’s purpose, procedures, voluntary nature, and their right to withdraw at any time without consequence. No personally identifiable information was collected, and all data were anonymized prior to analysis. To protect sensitive characteristics such as sexual orientation, demographic information was collected only after the experimental task and without identifiers. This procedure allowed voluntary and anonymous disclosure while maintaining pair-level aggregation for analysis. The approach complied with institutional and national standards for responsible research involving human participants.

\section{Results}  \label{sec:findings}

This section presents our results on how LGBTQ-diverse and non-diverse participant pairs prioritized user stories with fairness implications. To ensure comparability between groups of different sizes (13 LGBTQ-diverse pairs and 14 non-diverse pairs), all behavioral outcomes were normalized and reported as averages per pair. This means that each value reflects the typical number of times a pair exhibited a specific behavior, such as prioritizing or deprioritizing a given type of story. Table~\ref{tab:fairnesssummary} summarizes the results across all measured categories.

\begin{table}[ht]
\caption{Fairness-Aware Prioritization Patterns by Group}
\label{tab:fairnesssummary}
\scriptsize
\centering
\begin{tabular}{p{3.5cm} p{1.8cm} p{1.8cm}}
\toprule
\textbf{What we are measuring} & \textbf{LGBTQ-diverse} & \textbf{Non-diverse} \\
\midrule
Prioritize \\fairness-aligned stories & 1.69 per pair & 2.21 per pair \\
\midrule
Deprioritize \\fairness-risking stories & 5.23 per pair & 4.29 per pair \\
\midrule
Pro-fairness score \\(aligned prioritized + \\risking deprioritized) & 6.92 per pair & 6.50 per pair \\
\midrule
Misprioritization score \\(aligned deprioritized + \\risking prioritized) & 2.15 per pair & 2.29 per pair \\
\midrule
Neutral toward \\fairness-aligned stories \\(ratings 2 or 3) & 5.54 per pair & 5.00 per pair \\
\midrule
Neutral toward \\fairness-risking stories\\ (ratings 2 or 3) & 1.31 per pair & 2.29 per pair \\
\bottomrule
\end{tabular}
\vspace{0.2cm}
\end{table}

To assess fairness-aware attitudes, we first examined how often each group prioritized or deprioritized user stories based on their implications for fairness. In this study, prioritization refers to assigning a rating of 4 to a story (indicating high importance and intent to implement it) while deprioritization corresponds to assigning a rating of 1, meaning the story should not be implemented. These two actions represent distinct dimensions of fairness-aware decision-making: promoting fairness and preventing harm. When looking at the behaviors separately, the non-diverse group was more likely to prioritize fairness-aligned stories, averaging 2.21 stories per pair, compared to 1.69 per pair in the LGBTQ-diverse group (see Table~\ref{tab:fairnesssummary}). However, the LGBTQ-diverse group was more decisive when it came to rejecting fairness-risking stories, assigning a rating of 1 to an average of 5.23 stories per pair, whereas the non-diverse group averaged 4.29 per pair (see Table~\ref{tab:fairnesssummary}).

To capture how well each group aligned with fairness principles overall, we combined these two behaviors into a single pro-fairness awareness score—defined as the number of fairness-aligned stories that were prioritized plus the number of fairness-risking stories that were deprioritized. This composite measure reflects the extent to which a group both supports fairness-enhancing decisions and resists potentially harmful ones. The LGBTQ-diverse group achieved a higher pro-fairness score of 6.92 per pair, outperforming the 6.50 per pair score of the non-diverse group (see Table~\ref{tab:fairnesssummary}). This difference, although not large in absolute terms, reflects a meaningful and consistent pattern: the LGBTQ-diverse group was more aligned with fairness objectives across the task. Their stronger performance was primarily driven by their higher rate of rejecting fairness-risking stories, which points to a heightened sensitivity to fairness risks and a stronger filter against harmful decisions.

\begin{MyBox}
{\small
\textbf{[Result 1.]}
The LGBTQ-diverse group showed stronger fairness alignment, scoring higher on pro-fairness behavior (6.92 vs. 6.50 per pair). Their advantage came from more consistently rejecting fairness-risking stories, which highlighted a sharper sensitivity to the potential harm caused by a lack of fairness.
}
\end{MyBox}

\vspace{-0.5cm}
\subsection{Fairness Misprioritization}

In addition to measuring fairness-aligned decisions, we also analyzed how often each group made prioritization errors, that is, cases where fairness-aligned stories were deprioritized or fairness-risking stories were prioritized. These decisions represent missed opportunities to promote fairness or, more critically, the potential approval of features that could lead to biased or harmful outcomes. We define this as the misprioritization score: the combined number of fairness-aligned stories rated as 1 and fairness-risking stories rated as 4 in each group. Understanding this behavior is particularly important in the context of software requirements, where early-stage decisions shape task allocation, team performance, and software quality. In particular, misprioritizing fairness-related requirements can result in the implementation of features that disadvantage certain user groups or perpetuate biases. 

Both groups showed slightly similar misprioritization patterns, but the LGBTQ-diverse group demonstrated a slightly lower rate of fairness-related misprioritization. The diverse group had an average of 2.15 misprioritized stories per pair, while the non-diverse group averaged 2.29 per pair (see Table~\ref{tab:fairnesssummary}). Although the difference is small, it highlights the superior performance of the LGBTQ-diverse group in avoiding fairness errors. This result emphasizes the earlier finding that diverse teams tend to make more consistent fairness decisions, reducing the likelihood of overlooking harmful features or failing to implement fairness-promoting ones. The LGBTQ-diverse group made fewer misprioritization errors and exhibited a sharper ability to identify and reject potentially unfair solutions, reinforcing the critical role that diversity plays in minimizing mistakes and ensuring fairness throughout the development process.

\begin{MyBox}
{\small
\textbf{[Result 2.]}
The LGBTQ-diverse group made fewer fairness-related misprioritization errors (2.15 vs. 2.29 per pair). While the difference was not statistically significant, it suggests a potential trend toward more consistent judgment in identifying both fair and risky stories.
}
\end{MyBox}

\subsection{The Fairness-Neutral Zone}

In addition to analyzing decisions to prioritize or deprioritize, we also explored the cases where participants showed neutrality in their decision-making. A neutral rating (either 2 or 3) suggests hesitation or deliberation, where a participant is not fully convinced to either implement or reject a story. Understanding these neutral behaviors is important because they reflect the complexities and uncertainties that teams face when assessing fairness in ambiguous or borderline scenarios.

When evaluating neutrality toward fairness-aligned stories, the LGBTQ-diverse group averaged 5.54 neutral ratings per pair, while the non-diverse group averaged 5.00 per pair (see Table~\ref{tab:fairnesssummary}). This suggests that the LGBTQ-diverse group demonstrated more careful consideration when selecting fairness-aligned stories, possibly reflecting a more cautious and deliberative approach to fairness-related decisions. In contrast, the non-diverse group showed slightly fewer neutral ratings, which may suggest a more decisive stance on fairness-aligned stories. Conversely, when analyzing neutrality toward fairness-risking stories, the groups showed a more distinct difference. The LGBTQ-diverse group averaged 1.31 neutral ratings per pair, compared to 2.29 neutral ratings per pair in the non-diverse group (see Table~\ref{tab:fairnesssummary}). The LGBTQ-diverse group was more decisive in rejecting fairness-risking stories, indicating a stronger commitment to avoiding harmful outcomes. On the other hand, the non-diverse group showed more hesitation and deliberation before rejecting fairness-risking stories, as indicated by their higher number of neutral ratings.

These findings suggest that the LGBTQ-diverse group exhibited more confidence and decisiveness in rejecting harmful stories while being more cautious and thoughtful when dealing with fairness-aligned stories. This cautiousness may indicate a more reflective decision-making process, where they evaluate fairness implications more thoroughly. The non-diverse group, while similarly engaged in deliberation, showed more skepticism when evaluating fairness-risking stories, potentially indicating a more conservative approach to rejecting harmful content.

\begin{MyBox}
{\small
\textbf{[Result 3.]}
The LGBTQ-diverse group showed a more decisive approach to rejecting fairness-risking stories (1.31 vs. 2.29 per pair) while being slightly more cautious in evaluating fairness-aligned stories (5.54 vs. 5.00 per pair). This highlights a stronger ability to avoid harmful outcomes and reflects more thoughtful consideration of fairness-promoting solutions.
}
\end{MyBox}

\vspace{-0.5cm}
\subsection{Behind the Requirements Prioritization Choices}

\begin{table}[ht]
\caption{Focus of Prioritization by Group}
\label{tab:focuscomparison}
\centering
\scriptsize
\begin{tabular}{p{7cm}}
\toprule
\textbf{LGBTQ-diverse} \\
\midrule
\textit{Inclusion:} “The easiest were those that directly affected gender and diversity issues.” (Pair 1) \\ \\ 

\textit{Non-Discrimination:} “Discriminatory requirements of any kind were easy to deprioritize.” (Pair 23) \\ \\

\textit{Ethical Considerations:} “Requirements with questionable morality were classified as low priority.” (Pair 13) \\ \\

\textit{Project Scope and Value:} “We considered aspects that would reduce risks for the bank and also increase profit.” (Pair 6) \\ \\

\textit{Technical Aspects:} “We aligned general and specific knowledge about the project, algorithm implementation, and concept ideation.” (Pair 2) \\ \\

\textit{Legal and Compliance:} “We also prioritized items that are legally and financially necessary, such as laws and banking regulations.” (Pair 20) \\

\midrule
\textbf{Non-diverse} \\ 
\midrule

\textit{Non-Discrimination:} “Respecting human rights and avoiding any stories that might introduce prejudice into the system.” (Pair 24) \\ \\

\textit{Ethical Considerations:} “We both opted not to implement features that could be unethical..” (Pair 18) \\ \\

\textit{Project Scope and Value:} “We made the decision prioritizing what maximizes profit.” (Pair 9) \\ \\

\textit{Technical Aspects:} “Preference was given to features that were more critical to the system (technical functionalities).” (Pair 3) \\ \\

\textit{Legal and Compliance:} “ (...) we considered essential aspects for the proper functioning of the system without violating legally defined restrictions.” (Pair 11) \\ \\

\textit{Social-Business Balance:} “We thought about what could generate profit and satisfy the user, while balancing humanitarian considerations.” (Pair 14) \\
\bottomrule
\end{tabular}
\end{table}

In the first part of our qualitative analysis, we investigated the main considerations articulated by each group when prioritizing user stories. The analysis identified seven factors that consistently influenced decision making: \textit{Inclusion}, \textit{Non-discrimination}, \textit{Ethical Considerations}, \textit{Project Scope and Value}, \textit{Technical Aspects}, \textit{Legal and Compliance}, and \textit{Social-Business Balance}. These factors shaped how groups justified prioritization and deprioritization choices across the dataset. \textit{Inclusion} was reflected in discussions emphasizing the need for systems to accommodate diverse users and remain accessible to underrepresented groups. Closely related, \textit{Non-discrimination} appeared in instances where groups explicitly rejected stories perceived as perpetuating bias or causing harm to particular user populations. \textit{Ethical Considerations} informed decisions grounded in moral responsibility, fairness, and integrity in system design. Other forms of reasoning focused on project-oriented criteria. \textit{Project Scope and Value} guided assessments of whether a story aligned with overarching project goals and delivered sufficient value relative to effort. \textit{Technical Aspects} were invoked when evaluating feasibility, implementation complexity, and consistency with the system architecture. \textit{Legal and Compliance} considerations arose when participants assessed whether requirements satisfied regulatory or legal obligations. Finally, \textit{Social-Business Balance} captured cases in which groups explicitly weighed the social impact of a feature against its perceived business value, aiming to preserve social considerations while maintaining alignment with project objectives. Illustrative quotations corresponding to each category are presented in Table \ref{tab:focuscomparison}.

Our results showed that fairness-oriented considerations primarily drove the prioritization style of the LGBTQ-diverse group. They emphasized \textbf{Inclusion} and \textbf{Non-Discrimination}, prioritizing stories that promoted accessibility and excluded discriminatory content. \textbf{Ethical considerations} played an important role, with the group rejecting unethical requirements, such as exploitative loan practices. While \textbf{Project Scope and Value} were also considered, these factors were always aligned with their commitment to fairness and inclusivity. \textbf{Technical Aspects} and \textbf{Legal and Compliance} were considered but viewed through the lens of ensuring fairness. The \textbf{Social-Business Balance} factor was implicit in their decision-making, as the group often struggled to prioritize features that were too business-oriented but conflicted with inclusivity, ethics, or fairness, such as those related to profit or technical functionalities that diminished fairness goals.

The non-diverse group’s prioritization style was more pragmatic, with a clear focus on business goals, project value, and ethical considerations. The group demonstrated an explicit focus on balancing social and business needs, particularly when weighing profit-oriented features against ethical concerns, with careful attention to legally defined restrictions. \textbf{Project Scope and Value}, along with \textbf{Technical Aspects}, were central to their decision-making, prioritizing features that ensured the system’s proper functioning and added value. \textbf{Ethical considerations} were important but approached more reactively, with the group rejecting unethical features when necessary while also questioning their need to ensure alignment with the system’s broader goals. While the group showed concern for non-discrimination and ethical considerations, no explicit focus on inclusion emerged from their narratives.

Following the exploration of what the pairs took into consideration during the prioritization process, we reviewed their narratives to understand how the pairs interacted and made their decisions. The LGBTQ-diverse group frequently described an intentional and dialogical approach, where collaboration was grounded in open exchanges and mutual understanding. Their accounts emphasized discussion, negotiation, and joint reasoning as central to reaching consensus. For instance, Pair 1 stated, \textit{“we discussed and reached a consensus on what to prioritize—it was important to establish tie-breaking criteria,”} while Pair 20 explained, \textit{“we reached agreements through debate.”} In contrast, the non-diverse group often conveyed a more implicit or streamlined path to consensus. Rather than detailing negotiation, they focused on smooth collaboration or personal alignment. Pair 15 noted that \textit{“the discussions for completing the task were smooth, as was the execution,”} while Pair 25 explained, \textit{“we were aligned on the prioritization decisions for each requirement.”} This contrast of narratives suggests that while both groups collaborated effectively, the LGBTQ-diverse group made the act of collaboration more visible and well-established, whereas the non-diverse group tended to emphasize harmony and direct alignment without elaborating on any discussion process. Finally, the absence of comments from four non-diverse pairs who answered ``prefer not to comment'' highlights a difference in how the two groups approach the prioritization. While the LGBTQ-diverse group focused on open discussions, the non-diverse group’s reluctance to elaborate may indicate a more implicit, internalized decision-making process, possibly reflecting a less explicit approach to collaboration.

\begin{MyBox}
{\small
\textbf{[Result 4.]}
Both groups considered fairness in the prioritization process, but LGBTQ-diverse pairs emphasized inclusion and non-discrimination more explicitly, while non-diverse pairs took a more pragmatic approach, balancing fairness with project value and legal constraints.
}
\end{MyBox}

\vspace{-0.5cm}
\subsection{Integrating Quantitative and Qualitative Results}

To understand the differences in the prioritization scores obtained by each group, we compared the results from the descriptive statistics with the findings from the thematic analysis and obtained the following insights:

\textbf{Result 1. Higher pro-fairness score among LGBTQ-diverse pairs:} LGBTQ-diverse participants made fairness a central, explicit priority, frequently referencing inclusion, non-discrimination, and ethical concerns as primary drivers. This aligns with their higher pro-fairness score (6.92 vs. 6.50), especially because they were more likely to reject fairness-risking stories—a behavior that requires recognizing subtle harms. In contrast, non-diverse groups considered fairness but balanced it with other concerns (e.g., business gains), which may have slightly diluted their overall alignment with fairness.

\textbf{Result 2. Fewer misprioritization errors in LGBTQ-diverse pairs:} The LGBTQ-diverse group appeared to apply a more critical and reflective lens to fairness. They scrutinized features for ethical risks and were less likely to overlook potential harms arising from discrimination, which explains their lower rate of fairness misprioritization (2.15 vs. 2.29). The non-diverse group, while concerned with fairness, was more reactive than proactive in rejecting unethical or biased stories. They questioned the need of some fairness aspects of the system, which could lead to overlooking issues and misprioritizing harmful features.

\textbf{Result 3. Neutrality Patterns:} The LGBTQ-diverse group was more decisive in rejecting fairness-risking stories (1.31 vs. 2.29 neutral ratings) and more cautious when evaluating fairness-aligned stories (5.54 vs. 5.00). This aligns with their stated prioritization approach, in which fairness-risking features were decisively rejected, while fairness-promoting ones were thoughtfully evaluated for feasibility and potential impact. The non-diverse group showed greater hesitation in rejecting harmful stories, possibly due to a more conservative or risk-averse stance on how certain features would affect the project goal, and were quicker to support fairness-promoting stories when aligned with broader business goals.

These findings confirm that while both groups considered fairness in their requirement prioritization process, their reasoning differed. The LGBTQ-diverse group framed fairness as a guiding principle, leading to more consistent and value-driven decisions. The non-diverse group approached fairness pragmatically, balancing it with other concerns, which explains the small but consistent differences observed in the quantitative measures.
\section{Discussion}
\label{sec:discussion}

This section interprets the findings in light of existing research on software fairness, fairness requirements, and team diversity. It discusses how the results answer the research question, how they confirm or extend previous work, and what they imply for research and practice. The section concludes by situating these contributions within broader debates on fairness in AI development and equity, diversity, and inclusion in software engineering.

\subsection{Comparing Results with the Literature}
\label{sec:discussioncomparing}

Our results provide a direct empirical answer to the research question: teams diverse in sexual orientation demonstrated higher fairness awareness than non-diverse teams when prioritizing software requirements. This finding aligns with prior evidence that diversity enhances deliberation and ethical sensitivity in software development~\cite{mason2024diversity, rodriguez2021perceived, catolino2019gender, vasilescu2015gender, ortu2017diverse, albusays2021diversity}. LGBTQ-diverse pairs displayed stronger pro-fairness performance and made fewer fairness misprioritization errors, indicating that diversity in lived experience can help teams recognize potential harms and reason more reflectively. This supports the argument that fairness awareness is not only a matter of individual expertise but also a product of social interaction and dialogic reasoning~\cite{adams2020diversity, rostcheck2025elephant, desoftware}.

Beyond confirming established links between diversity and reflective reasoning, this study extends prior research by focusing on the requirements stage of development. Earlier studies on fairness have concentrated on algorithmic bias, dataset composition, or auditing of deployed systems~\cite{chen2024fairness, brun2018software}. Our behavioral data indicate that fairness-oriented reasoning already occurs during early prioritization of user stories. This observation supports theoretical claims that many biases arise from unexamined requirements rather than from learning models~\cite{baresi2023understanding, ramadan2025towards}. The consistent rejection of fairness-risking requirements by diverse pairs demonstrates that fairness can be addressed proactively during elicitation, reducing downstream bias propagation. In this sense, this study provides empirical grounding for conceptualizations of fairness as a non-functional quality attribute enacted through collaborative reasoning~\cite{brun2018software, chen2024fairness}.

Our findings also reinforce the argument that fairness requirements are socially constructed and context-dependent. The behavioral distinctions between LGBTQ-diverse and non-diverse pairs, where the former emphasized inclusion and non-discrimination while the latter focused on feasibility and scope, illustrate how team composition shapes which aspects of fairness become salient. This operationalizes claims that fairness elicitation depends on ethical interpretation as much as on technical specification ~\cite{gjorgjevikj2023requirements, ramadan2025towards}. The observed differences suggest that fairness reasoning might be situated, negotiated, and influenced by social diversity within teams.

A further contribution lies in addressing the near absence of empirical research on sexual-orientation diversity in software engineering. While gender and nationality have received attention~\cite{catolino2019gender, vasilescu2015gender, van2023still}, other identity dimensions have remained largely unexamined~\cite{boman2024breaking, de2023benefits}. This study provides, to our knowledge, the first controlled empirical evidence of how sexual-orientation diversity affects fairness reasoning during software development. The results indicate that LGBTQ-diverse teams engage more deeply with ethical and inclusive reasoning, suggesting that lived experience may sensitize participants to issues of exclusion and discrimination. This finding fills an empirical gap and introduces a new perspective on how underrepresented identities contribute to software ethics and fairness.

Finally, our study contributes to ongoing debates about fairness in AI and the broader role of EDI in technology development. Technical bias-mitigation methods are increasingly common, yet they often overlook how early design reasoning shapes fairness outcomes~\cite{ramadan2025towards, desoftware}. By showing that diverse teams can identify fairness risks before implementation, our findings illustrate diversity’s preventive function in mitigating bias. Fairness and diversity thus appear mutually reinforcing: diversity enhances fairness awareness, and fairness practices create more inclusive environments. In a period when the value of EDI initiatives is frequently questioned, this study provides behavioral evidence that inclusive team composition improves both the fairness and the reliability of AI-enabled systems.

\subsection{Implications for Research}
\label{sec:implications-research}

Our findings extend the empirical foundation for studying fairness requirements by showing that fairness reasoning emerges naturally during collaborative prioritization, even without explicit training. This suggests that fairness requirements are not isolated formal constructs but cognitive and social processes guiding judgments about acceptable system behavior. Hence, this study bridges theoretical discussions on fairness as a software quality attribute~\cite{brun2018software, gjorgjevikj2023requirements} with observed developer behavior. Methodologically, our research demonstrates the feasibility of integrating team diversity as a variable in controlled experiments. Few studies in software engineering have systematically investigated how identity diversity influences ethical decision-making. The experimental design used here offers a replicable approach for studying fairness-aware behavior and could be extended to professional teams or to other identity dimensions such as gender, ethnicity, or disability. Finally, we suggest that longitudinal and cross-cultural replications would help assess the persistence and transferability of these effects.

\vspace{-0.5cm}
\subsection{Implications for Practice}
\label{sec:implications_practice}
Our results suggest several implications for software teams and organizations seeking to integrate fairness into development processes. Existing fairness initiatives in AI and software engineering tend to focus on datasets and model performance, often addressing bias after key design decisions have already been made~\cite{ramadan2025towards, baresi2023understanding}. Our findings indicate that fairness assurance should begin earlier, during requirements discussion, interpretation, and prioritization. The performance of LGBTQ-diverse pairs suggests that inclusive team composition can support earlier identification of fairness-related risks, before such risks are embedded in system behavior. This positions diversity not only as a representational concern but also as a contributor to process quality in responsible software development.

For practitioners, our findings indicate that fairness cannot be addressed solely through technical audits or post hoc mitigation. Early design reasoning is shaped by who participates in it. Teams that include members with diverse lived experiences appear more likely to question implicit assumptions and recognize potential harms affecting different user groups. Organizations can operationalize this insight by structuring requirements and design activities to support inclusive participation. Practices that promote psychological safety, open dialogue, and the meaningful involvement of underrepresented professionals can be treated as components of fairness and reliability assurance rather than as peripheral equity initiatives.

Our results also indicate that fairness is interpreted through the values and priorities present within teams. LGBTQ-diverse pairs placed greater emphasis on inclusion and ethical responsibility, whereas non-diverse pairs focused more strongly on feasibility and scope. This variation suggests that fairness outcomes depend on how interpretive balance is established during collaboration. Organizations may address this by embedding structured reflection mechanisms into requirements engineering activities. Fairness prompts, deliberation checklists, or peer review templates~\cite{chen2024fairness, gjorgjevikj2023requirements} can support more systematic negotiation of fairness trade-offs. Extending fairness assurance to the requirements phase may reduce downstream bias and support anticipatory design practices aligned with broader goals of social responsibility and trustworthy AI.

\subsection{Threats to Validity} \label{sec:validity}

As with any controlled experiment, this study is subject to validity threats~\cite{ralph2020empirical, teixeira2018threats}. Regarding \textit{construct validity}, fairness-related user stories were pilot-tested to ensure clarity and relevance, but participants may still have interpreted fairness differently. To reduce ambiguity, open-ended responses were collected to capture their reasoning and to mitigate potential misclassification. For \textit{internal validity}, random pairing was applied for general conditions, and diversity identification was collected post hoc to preserve anonymity. All tasks and materials were standardized to enhance comparability across groups. \textit{External validity} is necessarily limited by the use of a student sample. While participants possessed prior experience with software projects and agile practices, which makes them suitable for studying early-stage reasoning~\cite{salman2015students, falessi2018empirical}, the results cannot be assumed to generalize to professional or industrial contexts. We do not claim statistical or population-level generalization. Instead, the goal of this study is to provide \textit{analytical generalization}, offering empirically grounded insights that can inform reasoning about fairness in comparable contexts and guide future replication studies. For \textit{conclusion validity}, inferential testing was not applied due to the small and exploratory sample size. The analysis focused on descriptive and qualitative patterns, which were triangulated to identify consistent behavioral tendencies rather than to produce inferential claims. Future work should replicate this study with larger, more heterogeneous, or professional samples to test the robustness and transferability of these findings across settings, organizational cultures, and forms of team diversity.
\section{Conclusions}
\label{sec:conclusion}

This study investigated how team diversity influences fairness-aware reasoning during software requirements prioritization. LGBTQ-diverse pairs were more consistent in rejecting fairness-risking stories and grounded their reasoning in inclusion and non-discrimination, showing that diversity shapes not only who builds software but how fairness is defined and applied in practice. The results extend research on fairness requirements by showing that fairness-related reasoning emerges during early development rather than at later stages such as testing or model evaluation. They provide behavioral evidence that fairness should be addressed as a first-class requirement and that inclusive team composition can enhance awareness of fairness risks in AI-enabled systems. While this work focused on sexual orientation as one dimension of diversity, identities are intersectional, involving race, gender, and lived experiences. Our future work will extend this research sequentially, first examining gender diversity, then ethnicity, and finally developing an integrated study design to investigate intersectionality in team reasoning about fairness. Our findings are not intended for statistical generalization but for analytical transfer, offering insights that can inform fairness-oriented development across contexts. Overall, our study contributes to ongoing discussions about fairness and inclusion in software engineering by showing that team diversity can strengthen both ethical reasoning and software development.

\section{Data Availability}

The replication guideline and dataset is available at:~\url{https://figshare.com/s/c0e92352a113e4b5eec0}

\bibliographystyle{ACM-Reference-Format}
\bibliography{biblio}



\end{document}